# Precision bond lengths for Rydberg Matter clusters $K_N$ (*N* = 19, 37, 61 and 91) in excitation levels *n* = 4 - 8 from rotational radio-frequency emission spectra

Leif Holmlid

Atmospheric Science, Department of Chemistry, Göteborg University, SE-412 96 Göteborg, Sweden

Email: holmlid@chem.gu.se. Phone: +46-31-7722832. Fax: +46-31-7723107.

Abstract

Clusters of the electronically excited condensed matter Rydberg Matter (RM) are planar and six-fold symmetric with magic numbers *N* = 7, 19, 37, 61 and 91. The bond distances in the clusters are known with a precision of ± 5% both from theory and Coulomb explosion experiments. Long series of up to 40 consecutive lines from rotational transitions in such clusters are now observed in emission in the radio-frequency range 7-90 MHz. The clusters are produced in five different vacuum chambers equipped with RM emitters. The most prominent series with *B* = 0.9292 ± 0.0001 MHz agrees accurately with expectation (within 2%) for the planar six-fold symmetric cluster $K_{19}$ in excitation level *n* = 4. Other long series agree even better with $K_{19}$ at *n* = 5 and 6. The ratio between the interatomic distance and the theoretical electron orbit radius (the dimensional ratio) for $K_{19}$ in *n* = 4 is found to be 2.8470 ± 0.0003. For clusters $K_{19}$ (*n* = 6) and $K_{37}$ (*n* = 7 and 8) the dimensional ratio 2.90 is the highest value that is found, which happens to be exactly the theoretical value. Clusters $K_{61}$ and $K_{91}$ in *n* = 5 and 6 have slightly lower dimensional ratios. This is expected since the edge effects are smaller. Intensity alternations are observed of approximately 7:3. The nuclear spins interact strongly with the magnetic field from the orbiting electrons. Spin transitions are observed with energy differences corresponding accurately (within 0.6%) to transitions with apparent total $\Delta F = -3$ at excitation levels $n = 5$ and 6. The angular momentum coupling schemes in the clusters are complex but well understood.

# 1. Introduction

Planar Rydberg Matter (RM) clusters have previously been studied by laser induced fragmentation of RM clouds in high vacuum [1-5]. They are six-fold symmetric with so called magic numbers $N$ = 7, 19, 37, 61 and 91. Using pulsed laser induced Coulomb explosions, bond lengths and cluster sizes have been determined in several different cases, confirming theoretical predictions of bond lengths and electronic state [6]. In these calculations, a quasi-classical approach with electron correlation was used to determine the dimensional ratio $d/r_n$ = 2.9, where $d$ is the distance between the ion cores in the RM (the bond distance) and $r_n$ is the Rydberg electron orbit radius for principal quantum number (excitation level) $n$. The Coulomb explosion method has given bond lengths with a precision as good as ± 5% in agreement with theory [7]. However, it is desirable to determine the bond distances with higher precision.

A more precise method to measure the bond lengths in the RM clusters should be rotational spectroscopy of the clusters. There are good reasons to expect that emission spectra can be observed in the experimental setups where RM clusters are produced, especially since efficient stimulated emission from RM in the IR is observed [8], even in the form of the RM laser [9,10]. Calculations of the rotational constants for the commonly observed planar RM clusters $K_{19}$ and $K_{37}$ in excitation levels $n$ = 4 – 6 show that rotational spectra should be observed in the RF range below 100 MHz. These values of $n$ in RM clusters are regularly found also in other types of experiments like the laser induced fragmentation studies. Long series of equally spaced lines are indeed observed in emission in the experiments. Several rotational spacings are found repeatedly, and rotational constants can be found in close agreement, within 1%, with the previously observed and theoretically predicted cluster forms.

Here, we report on the use of rotational spectroscopy in the radio frequency range to determine cluster bond distances in nine different types of RM clusters. For example, in $K_{19}$ ($n$ = 4) the bonding distance is determined with a precision of $1 \times 10^{-4}$, providing a value of the dimensional ratio $d/r_n$ = 2.8470 ± 0.0003. These high-precision measurements demonstrate yet another method that gives conclusive evidence of Rydberg Matter from the agreement between experimental results and theory. From studies of the other cluster forms with $n$ = 4-8 and $N$ = 19, 37, 61 and 91 important trends are found in the dimensional ratio with $n$ and $N$.

## 2. Theoretical

### 2.1. RM clusters and their radiative transitions

RM was predicted by Manykin et al. in 1980 [11], and studied theoretically by his group [12,13]. The first experimental study of RM identified RM as a condensed material with extremely low work function and high conductance in a cesium plasma diode [14]. This work has been confirmed independently [15]: these recent authors report even on the size of the RM clusters in their experiments. Independent observations and suggestions of RM in other types of plasma systems have also been published [16,17]. Independent studies of the RM emitter materials used here and of similar catalysts show that Rydberg species are formed at their surfaces [18-20]; this is the first step in formation of RM clusters.

The RM clusters are planar, mainly sixfold symmetric, usually with magic numbers 7, 19, 37, 61 and 91 [2]. The atoms forming the clusters at the surface of the RM emitter are interacting long-lived circular Rydberg species with the outer (valence) electrons in almost circular ($l$ = $n$-1) orbits. The bonding in the RM clusters is almost metallic, with the electrons delocalized in a conduction band but still with high *l* values. Classical field calculations have been done, with electron correlation taken into account in an approximate fashion [6]. These calculations show that a binding in the RM phase can only exist if all electrons are coherent and thus all valence electrons have the same excitation level characterized by the initial principal quantum number *n*. The distance between the atoms in RM is found to be 2.9 $n^2$ $a_0$, where $a_0$ is the Bohr radius and *n* is the principal quantum number (excitation level). In the case of K atoms as in the present case, the lowest possible excitation level is *n* = 4. The corresponding Rydberg configuration for the separate atoms is in this case $3p^6$ $4f$, thus a $^2$F state. The electrons in the RM clusters still have large orbital angular momenta *l*, which means that they give rise to strong magnetic dipoles, with each electron orbit giving a contribution of *m* = (*n*-1) $\mu_B$ [21], with $\mu_B$ being the Bohr magneton. Thus, the clusters studied here in general have a magnetic dipole two orders of magnitude larger than most molecules. Experimental [8,9,22,23] and theoretical evidence in the case of RM strongly indicates that no typical quantized vibrations exist in RM. This is so, since the atoms (core ions) are kept in place by the simultaneous slow electronic motion. A separation between electronic energy and ion kinetic energy is not possible since the velocities of the electrons and ions are of similar size, which implies that the Born-Oppenheimer approximation is not valid. This means that the electronic and nuclear motions can not be separated: indeed, nuclear vibrations are part of the electronic orbiting motion.

Energetically, just a few types of transitions are possible in the spectral range studied. The largest *n* value in the clusters studied is *n* = 8. An electronic transition *n* = 9 ? 8 in RM for one electron corresponds to an energy of 360 cm$^{-1}$ or 1x10$^{13}$ Hz, much larger than the maximum frequency of 100 MHz observed here. Other possible *n* transitions correspond to even larger energies. Only for *n* > 400 do the transitions fall into the studied spectral range, but such clusters will probably not be able to exist anywhere in the universe [24]. Thus, the observed transitions can not involve changes in orbital angular momentum of the electrons in the clusters. A flip of an electron spin corresponds to approximately $1.8 \times 10^{11}\ n^{-5}$ Hz, or 6 MHz at $n = 8$, thus within the range of observed frequencies. This is found from the spin-orbit coupling energy. The frequency corresponding to a spin-flip of an electron at $n = 4$ is of the order of 190 MHz, thus too large to be observed here. Thus only spin flips at $n = 5 - 8$ are within the observed frequency range. Another degree of freedom is the energy of electron translation in the RM clusters, which probably has quanta of the right size to be observed in the experiments [24]. Agreement is however not found at the excitation levels $n = 4 - 8$ observed here. This leaves the main processes observed as rotational transitions of the clusters and nuclear spin transitions, as described further below.

The formulae for the rotational energy levels in the oblate symmetric rotors are found in standard texts [25]. The moment of inertia that determines the pure rotational transitions is the intermediate moment $I_B$. In the case of an oblate planar molecule with $I_A = I_B$ due to symmetry, the moments of inertia are related as $I_C = 2I_B$. The moments of inertia $I_B$ for the planar six-fold symmetric tops are easily found, assuming that the bond distances $d$ are constant in each cluster. They can be summarized concisely as $I_B/(m_K d^2) = 24$ ($N = 19$), 93 ($N = 37$), 255 ($N = 61$) and 570 ($N = 91$).

The isotopic composition of K is 93.9% $^{39}$K and 6.7% $^{41}$K. The smallest cluster $K_{19}$ studied here contains one $^{41}$K nucleus on average, and the larger clusters contain two or more heavy nuclei per cluster. This will of course influence the locations of the lines. With one or two heavy nuclei, the clusters have several different forms with their special rotational constants. The usual description [25] of the slightly asymmetric clusters shows that the rotational constant *B* is replaced by the average ½(*A*+*B*), where *A* is very close to *B*. In the most easily analyzed case of $K_{19}$, the extreme position of the $^{41}K$ atom is at one of the apexes of the cluster, with a probability of 0.32. In such a case, $I_B = 24 \times 39u\, d^2$ while $I_A = 24 \times 39.33u\, d^2$. At a frequency of 50 MHz, this means a shift of 430 kHz. Such a splitting is easily detected but not easily identified due to the large number of small intensity lines often observed. Another factor that is decisive for the detection is that a high density of photons will give stimulated emission and amplification of the line. Thus, preponderance for the stronger transitions due to clusters containing only $^{39}K$ is expected, at least for the smaller clusters like $K_{19}$. All calculations here accordingly use 39 *u* as the mass of the K atoms.

## 2.2. Resolution and stimulated emission

The rotational signals for $K_{19}$ clusters are narrow lines, with an observed half-width down to 20 Hz, determined with a limit of 10 Hz of the resolution bandwidth in the spectrum analyzer. For larger clusters like $K_{37}$, the lines are broader, of the order of a few kHz and they often show an unresolved structure. Due to the large mass even of the $K_{19}$ clusters, which are the smallest clusters observed, the Doppler width is expected to be 10-20 Hz at $T$ = 100-300 K, with the low temperatures reached by self-cooling, mainly by IR emission [8-10].

The RM clusters form a cloud in the vacuum chambers, with a typical size of a few cm, as observed directly from laser probing of the cloud [3,4]. In this cloud, the interaction is weak enough that different excitation levels *n* are possible in different clusters, but it means that the clusters do not move as free gas molecules. Self-cooling of the cloud primarily by stimulated emission takes place down to kinetic temperatures of 20 K as observed in several experiments [4]. Stimulated emission is observed in the IR in the range 800 – 16000 nm, and is the basis for the broadly tunable RM laser [8,9]. Since the ratio of the Einstein coefficients $A_{spon}/B_{stim}$ varies as $\nu^3$, stimulated emission is relatively much faster than spontaneous emission at low frequencies. In the present case, $A_{spon}/B_{stim} = 1.7 \times 10^{-35}$ J m$^{-3}$ Hz$^{-1}$. Using a bandwidth of 10 Hz this corresponds to a very low photon density of $6 \times 10^{-9}$ m$^{-3}$ at the point where stimulated emission is as fast as spontaneous emission. The lowest signal level included, 1 μV in 50 Ω, corresponds to $> 10^{11}$ photons s$^{-1}$ or to $> 10^3$ photons m$^{-3}$. Thus, it is apparent that stimulated emission dominates strongly over spontaneous emission in the experiments. Since stimulated emission is efficient, also absorption must exist in the RM in the experiments described here. The balance between self-absorption and emission is not known. In general, RM is dark, with no strong spontaneous emission observed in the visible or the IR out to 16 000 nm. In Fig. 1, the different states involved in the stimulated emission process are shown. The RM clusters are formed by thermal desorption from the emitter, in excitation levels $n = 20$ -30 [9,26]. The high excitation level of the clusters is decreased by efficient stimulated emission in the IR [8,9], and the cluster rotation is cooled down by stimulated emission in the RF as shown in the figure. The rotational state of the clusters is described further in the next subsection.

**2.3. Angular momentum couplings**

The RM clusters have angular momenta due to overall rotation $\underline{R}$, electron orbiting motion $\underline{L}$, spin $\underline{S}$, and nuclear spin $\underline{I}$. They are symmetric oblate rotors. Since the magnetic field from the electronic orbiting is large at least at low $n$ values (see below) and directed along the figure axis of the cluster, it is possible that several other momenta like the spins $\underline{S}$ and $\underline{I}$ couple to the figure axis of the cluster. The general configuration is shown in Fig. 2, where nuclear spin angular momentum is not included. Note that the $\underline{S}$ vector will in general be in an opposite direction relative to $\underline{L}$, since this is the lowest energy state. This coupling scheme is similar to Hund's case $a_\alpha$ in the case of a diatomic molecule [25], but the cluster here is not linear but planar. Further, $\underline{L}$ is fixed along the figure axis in the classical limit, which means that $M_L = L$. Since $L$ is large, equal to 19×4 = 76 as the minimum for the clusters studied here, this means that the vector $\underline{L}$ is very close to the figure axis also in quantum mechanical terms. The same will often be true for the spin vector: with a relatively small value down to $S = 9.5$ here, the vector $\underline{S}$ is probably close to the figure axis. The individual spins of the electrons may still have different directions, giving a total $\underline{S}$ vector coupled to and precessing around the figure axis but not locked to it. The rotation of the cluster is that of a symmetric rotor, with the angular momentum vector $\underline{R}$ precessing around the direction of $\underline{J}$. The component along the figure axis, indicated with $\underline{K_c}$ in the figure, is constant in a pure rotational radiative transition, while the perpendicular component $K_b$ may change in a transition. A rotational transition in the RM cluster will in the case in Fig. 2 involve a transition between two levels with large values of $J$, where $J$ normally is not smaller than $L - S$.

The clusters studied are not in thermal equilibrium, but in an excited state and in fact an inverted state due to their origin in deexcitation of clusters in higher electronic levels. When a cluster is deexcited from its initial intermediate value of $n = 10 - 20$ down to the levels observed here, i.e. $n = 4 - 8$, the orbital angular momentum is partially removed by the emitted photon or photons. Some of the angular momentum may also be absorbed into overall rotation, so when the electrons deexcite the cluster rotation is increased. This is an internal process in the cluster and does not involve any photon formation. In such a case Fig. 2 is not meaningful, since the R vector will be directed along or very close to the figure axis. The clusters generally survive in the vacuum chamber after deexcitation and are re-excited by absorption of thermal photons, as shown in the temperature variation experiments performed in the RM laser setup [9,10]. The photons are absorbed from random directions and with random polarization during the rotation of the cluster. Such clusters are long-lived, with lifetimes probably of the order of minutes, and can reach a state where they interact minimally with the surroundings, both in collisions and in radiation. This form of cluster is thus the most general one that is observed (or not so easily observed) in the experiments. The strong IR and RF emissions observed in the experiments are signs of the process to a minimal interaction state. This situation is shown in Fig. 3, where in the left-hand panel the overall rotation angular momentum is directed approximately counteracting the $\underline{W}$ vector, leading to small and intermediate $J$ values. The $J$ number will generally be half integer, since the number of electrons is odd. Note that if $\underline{R}$ is exactly anti-parallel to $\underline{W}$, the vector $\underline{J}$ is also along the figure axis of the cluster and there is no time-variable magnetic dipole and no interaction with the radiation field. Thus, the cluster in this minimal interaction form is entirely dark. It may be expected that the excitation level of this dark form in the vacuum chambers is $n = 4$; this is the lowest possible value for K RM. Due to collisions, the clusters will transfer to the form shown to the right in Fig. 3, where the $J$ number still is low but rotational transitions are possible due to the rotating magnetic dipole. This is probably the form detected for $K_{19}$ ($n = 4$), but it is less likely for higher $n$, as described in the discussion section.

Larger clusters may have the slightly more general coupling form of Fig. 4 since collisions are more likely and since they may have formed by relatively recent mergers of smaller clusters, giving larger rotational angular momentum $R$. Most such interaction processes will tend to rotate the cluster around an $I_B$ axis. In Fig. 4, the value of $K_c$ is thus shown to be small and the vector $\underline{R}$ is located close to the cluster plane. In this figure also the weaker coupling between $\underline{L}$ and $\underline{S}$ at larger values of $n$ and thus weaker magnetic fields due to the electron orbiting (see below) is taken into account, just showing the final vector $\underline{W}$. The quantum number $W$ may in this case have integer values due to the direct coupling between $\underline{l}$ and $\underline{s}$ in each atom when the field is weaker.

### 2.4. Internal magnetic field in RM

The orbiting electrons in an RM cluster give rise to a strong magnetic field. The most important aspect of this is the interaction of the $I = 3/2$ nuclear spins in the $^{39}$K and $^{41}$K nuclei with the magnetic field. The magnetic field strength in the center of a current loop (i.e. the electron orbit) is along the central axis and has the value

$$B = \frac{\mu_0}{4\pi} \frac{2IA}{r_n^3} \tag{1}$$

where $I$ is the current in the loop, $A$ is the area of the loop and $r_n$ is its radius. This radius varies with the principal quantum number $n$ for the orbiting electron. Introducing the area $A = \pi r_n^2$, the current $I = ev_n/(2\pi r_n)$ with $v_n$ the velocity of the orbiting electron, and standard expressions for $r_n$ and $v_n$ give

$$B = \frac{\pi \mu_0 e^7 m_e^2}{8 \epsilon_0^3 h^5} \frac{1}{n^5}. \tag{2}$$

Evaluating this gives

$$B = 12.508 \times n^{-5} \text{ T} \tag{3}$$

for the magnetic field strength at the center of the current loop, thus at the nucleus. The number $n$ is the principal quantum number (level of excitation) for the RM electron, and the magnetic field decreases rapidly for higher $n$. Values of $B$ for the $n$ values of interest here are given in Table 1.

In the RM cluster, many electrons exist, each one orbiting one nucleus in the classical limit as shown in the quasi-classical calculations in [6]. This means that the magnetic field from the surrounding atoms will influence the field strength at each nucleus. In fact, since the magnetic dipoles of the electronic motion are all in the same direction, the surrounding dipoles will slightly decrease the field at each nucleus. Using Eq. (1) and its corresponding form for the parallel magnetic field strength from neighbour group $i$ at a general distance $r_i$ from the dipole

$$B_i = \frac{\mu_0}{4\pi} \frac{IA}{r_i^3} \qquad (4)$$

gives the contribution from two consecutive surrounding (planar) neighbour shells as

$$B_{sum} = \frac{\mu_0}{4\pi} \frac{2IA}{r_n^3} \left(1 - \frac{3}{2.9^3} - \frac{6}{(2\times 2.9)^3}\right). \qquad (5)$$

Here, the distance between the atoms (dipoles) has been set to exactly 2.9 $r_n$ [6] (see further below). The expression in parantheses is equal to 1 - 0.123 - 0.031 = 0.846. Thus, using this correction for the first and second circles of neighbours gives a more accurate value for the constant in Eq. (3) as

$$B = 10.582 \times n^{-5} \text{ T} \qquad . \qquad (6)$$

Of course, not all atoms in an RM cluster have so many neighbours, and the edge nuclei in a cluster will experience a larger magnetic field, maybe 10% higher than that for the central nuclei in the cluster.

## 2.5. Nuclear spin transitions in RM clusters

The component of nuclear magnetic dipole moment in the direction of the magnetic field is

$$m_z = \frac{1}{\hbar} g_I \mu_N m_I \qquad (7)$$

where $g_I$ is the g-factor of the nucleus, $m_I$ ranges between I and –I of the nuclear spin quantum number, and $\mu_N$ is the nuclear magneton. In the case of both nuclei $^{39}$K and $^{41}$K the value of I is 3/2. This means that $m_I$ = 3/2, ½, -1/2, and -3/2 are the values possible for each K atom. The frequency splitting between two levels with different $m_I$ using the selection rule $\Delta m_I = \pm 1$ is

$$\Delta \nu = \frac{1}{\hbar} g_I \mu_N B \qquad (8)$$

With $g_I(^{39}K) = 0.260977$ and $\mu_N = 5.0508 \times 10^{-27}$ Am$^2$ this gives

$$\Delta \nu \, n^5 = 1.3230 \times 10^8 \text{ Hz.} \qquad (9)$$

Excitation level $n = 5$ gives 42.34 kHz, and $n = 6$ gives 17.01 kHz for each nucleus making a transition $\Delta m_I = \pm 1$. The formulas above may be expressed also by using the magnetic dipole interaction constant $a$ [27] which is

$$a = -g_I \mu_N B \qquad (10)$$

with the value of $B$ for the internal field from Eq. (6). The resulting splitting of the energy levels follows the Landé interval rule [27]. Nuclear spin $\underline{I}$ adds to $\underline{J}$ in the cluster with the result as the vector $\underline{F} = \underline{I} + \underline{J}$. Using the quantum number $F$, the energy differences are $\Delta E = aF$, where $F$ is for the high $F$ level (upper level for a "normal" structure). This means that the transitions are equally spaced with $\Delta \nu = a/h$, as described above, with each line corresponding to one $F$ value.

## 3. Experimental

### 3.1. RM emitters

Most of the five machines used have been described previously: Chamber 1 is the RM laser equipment [8-10]; Chamber 2 has a Mo emitter foil mounted in front of a K molecular beam source (cold in the present case); Chamber 3 is the UHV chamber where laser fragmentation experiments are performed [5,7]; Chamber 4 has an RM emitter mounted for micro-spectroscopic studies [28]; Chamber 5 is similar to Chamber 4, but used for single-mode IR laser studies of RM [22,23]. The experiments are performed in a vacuum between $10^{-4}$ and $10^{-9}$ mbar, at sample temperatures between 300 and 900 K. The RM emitter material used in all of them but Chamber 2 is a K promoted iron oxide catalyst similar to Shell S105, used for production of styrene from ethyl benzene [29]. Other alkali doped catalyst types also give RM clusters in desorption. The heating of the catalyst is by an AC current through a metal foil tube or boat holding the small emitter samples. Two typical catalyst emitter holders are shown in Figs. 5 (Chamber 1) and 6 (Chamber 4 and 5). In Fig. 6, a few pieces of the fresh iron oxide catalyst are also shown. Most of the K atoms in the emitter material (probably only the ionically bound K atoms) have to be removed by heating prior to experiments. It seems that the $K^+$ current observed from the heated emitter with an accelerating field strength < 10 V/cm has to be brought to a low value to bring the sample into operating conditions. In general, the operating time needed before a fresh emitter sample will function well is days to months under actual low-pressure conditions, depending on the residual gas composition in the vacuum chamber, especially the relation between oxygen and hydrocarbon partial pressures. Hydrocarbon molecules reaching the catalyst will decompose and finally form graphite on the surface, since the catalyst is designed to strongly increase the rate of hydrogenation and dehydrogenation reactions. Admission of air to a pressure of $10^{-5} - 10^{-4}$ mbar in the vacuum chamber is often required to reduce the rate of graphite formation on the catalyst surface, if the hydrocarbon partial pressure is high in the chamber: a too thick graphite layer will strongly decrease the rate of K diffusion from the bulk. However, in most cases the procedure can be varied considerably with no great consequences for the final emitter performance. Depending on the status of the iron oxide emitter material after initial preparation, especially the degree of depletion of K in the material and the amount of graphite coverage on the surface (for example as observed in the center of Fig. 6), the heating cycle needed in each experiment (temperature and time required) will vary before a dense enough RM cloud will be formed in the apparatus. Extensive operating experience is required to optimize both the initial emitter preparation process and the actual RM formation process under the actual residual gas conditions: the RF emission described here appears to be the most convenient device discovered so far to observe the RM formation. The most prominent and intense rotational series described below is not observed at all in Chamber 2 with its plain foil emitter. In the other chambers, this series is observed as expected after heating the RM emitter for several hours. This is the same type of procedure as used for forming RM in the regular experiments run in these chambers.

### 3.2. Interference by external sources

The spectrum analyzer used in the experiments is Hewlett-Packard 8590B or 8596E. The studies reported have been performed in five different types of vacuum chambers, standing in two adjacent laboratories with a distance between each of them of typically five meters. The typical dimension of the chambers is 0.3 – 1.0 m. The studies are made at radio frequencies in emission from the RM clusters, at frequencies of 15-90 MHz, i.e. with wavelengths of 3-20 m, generally larger than the distance between the different chambers. This means that similar signals would be observed in all the chambers, if the signals were of external origin. A possible problem in the experiments could be the observation of external radio sources in the building and outside the building, for example used for broadcasting purposes. In the laboratory, the damping due to the reinforced concrete in the building is considerable, decreasing the signal at a broadcasting FM frequency of 89 MHz by up to a factor of four. This brings the strongest broadcasting signals to the same or slightly higher levels than the signals from the RM clusters. The sharp signal lines are easily distinguished from broad FM radio signals. Pulsed narrowband AM radio signals will usually appear at very low levels due to the averaging over 100 – 250 s used. The main feature used here to identify the rotational spectral series for the small clusters $K_{19}$ is that they belong to long series of equally spaced lines extending over 20 – 40 MHz.

No substantial change but in signal size is observed for different antenna constructions. Various types have been tested, straight or bent 10 cm pins into the vacuum chamber, discs with a diameter of a few cm in the vacuum, or the entire chamber wall. The RM emitters are at temperatures of 300-900 K during the experiments, while most of the RM cloud formed is at lower temperature, 100 K or lower due to self-cooling by stimulated emission. Another aspect of the setup is if the clusters can exist undisturbed, i.e. without collisions, for long enough times to give the sharp line-widths observed. Several factors indicate this: the clusters exist at low temperature in the RM cloud, the antenna monitors the field in the entire chamber due to the long wavelength, and stimulated emission dominates.

It is first necessary to demonstrate that the spectra are not due to any kind of external source. In Fig. 7, typical spectra of frequencies at the five chambers are shown. In general, all lines with a peak level above 1 $\mu$V at a bandwidth of 3 kHz were recorded. The scan width in each RF scan is typically 2 MHz, and each spectrum in Fig. 7 is thus composed of 40 different scans. The relative intensities in each spectrum are not the same as if the spectrum was taken at once, since the distribution of clusters in the RM cloud changes with time and running conditions, in different ways in the different chambers: this is expected since all the different emitters have different histories of usage. If RF signals from external sources outside the vacuum chambers were of importance, similar signals would have been observed at more than one machine due to the long wavelength of the radiation. The spectra are generally very dissimilar, and they also vary with several experimental parameters. The one entitled "Chamber 3 RM run" is measured in an experiment during which a signal was produced for a few hours after prolonged heating of the RM emitter and subsequent cooling. The spectrum shown for Chamber 2 is only found with the foil emitter at high temperature; without emitter heating, no lines are observed. In Chamber 5, a rotational spectrum can only be observed after heating the emitter to the same temperature as used in the previous RM experiments in this chamber [22,23]. Another observation is that the signal intensity at the internal antenna changes if the conductors (feedthroughs) connected to the emitter in Chamber 1 are grounded or floating, which shows directly that the source of the radiation is at the emitter. Since the spectra are influenced by several experimental parameters, the source is in the chamber in each case: a direct search for the source of RF radiation with a short antenna gives the same result.

4. Results

Many series of lines exist in the spectra. In the experiments, no prior assumptions have been made concerning the nature of the signals. Two different possibilities to study the spectra have presented themselves, a) the study of the most intense series with constant large spacings between the lines, also displaying very long series of lines extending over several tens of MHz, and b) the study of the series in bands with small spacings, relatively few lines in each series and often intensity alternations between consecutive lines. Both these possibilities have been used, as shown in sections 4.1 and 4.2 below. There also exist intermediate types of series, with intermediate spacings and intermediate lengths, but the identification of most of them requires that the other series are well known since they usually do not appear alone under tested emitter operating conditions. Thus, all series that can be observed are still not interpreted in detail, and the possibility that asymmetric rotors contribute to the spectra can not yet be rejected. Certainly, such lines from isotopically complex clusters can be observed. The spectral series observed are however so far systematized in terms of the symmetric oblate RM clusters, providing excellent agreement with the theoretical predictions. In Table 2, the spacings of the type 2*B* are shown for most observed series; the not yet observed spacings (within parantheses) are calculated from RM cluster theory. The table is limited to the commonly observed RM cluster sizes $N$ = 19, 37, 61 and 91 and the most commonly observed excitation levels at $n$ = 4 – 8 for K RM clusters. The observed rotational constants are also used to determine the corresponding dimensional ratios $d/r_n$ in Table 3, displaying important trends in this ratio. In all calculations, the mass of each K atom is set equal to 39 mass units, as motivated in the theoretical section.

### 4.1. Long series due to $K_{19}$ clusters

Under many conditions, one long and intense series of lines dominates the spectra. This series can be observed directly in several of the plots in Fig. 7. A typical plot of 30 line positions is shown in Fig. 8 providing an excellent linear fit. This series is measured in one chamber during one day to eliminate drifts in the RM cloud and in the electronics, but similar results are found repeatedly, also in other chambers as can be seen already in Fig. 7. The fit parameters are given in the legend. The difference between the points 2***B*** is 1.8583 ± 0.0003 (standard error) MHz. Repeated measurements in two different chambers agree within error limits and give a more precise value for this rotational spectrum of $2B = 1.8585 \pm 0.0002$ MHz, a rotational constant $B = 0.9292 \pm 0.0001$ MHz and a moment of inertia $I = 9.0315 \times 10^{-42}$ kg m$^2$. The values are also given in Table 4. Due to the excellent linear fit with equally spaced lines, the cluster is a symmetric top (or possibly a linear molecule, excluded due to theoretical reasons): the $K$ quantum number selection rule for a symmetric top is $\Delta K = 0$. Rather small centrifugal stretching is expected since the rotational velocity is low with a rotational frequency of the order of 1 MHz: see further in the discussion. An RM cluster $K_{19}$ gives excellent agreement: a bond distance of 2.4105 nm is found, which agrees with the approximate theoretical form 2.9 $a_0 n^2$ [3,4,6] with $n = 4$ within 2%. Expressed differently, it means that the dimensional ratio $d/r_n$ is now determined with high precision, giving a value of $2.8470 \pm 0.0003$ for $K_{19}$ ($n = 4$). This value is included in Table 3. The results are also collected in Table 4.

Using new emitter samples at high temperature, thus with a low coverage of carbon on the surface but probably with rapid desorption of RM clusters into higher excitation levels, the results in Fig. 9 are found. The line with smallest slope shows the behaviour for $K_{19}$ clusters at $n = 6$ at the highest emitter temperature, while the upper line shows the plot for $K_{19}$ clusters at $n = 5$ at somewhat lower temperature. It is not possible to specify in advance the conditions for forming the specific well defined cluster forms; instead, the analysis of many runs identified some where almost all lines fall into the pattern of one specific cluster type. That some of the transitions are not observed in the plots is partly due to overlap with other bands containing series due to heavier clusters with $N = 37, 61$ and 91. See further Tables 3 and 4 for comparisons between the accurate rotational constants and bond distances derived. The values of $2B$ are included in Table 2, and the resulting dimensional ratios $d/r_n$ are shown in Table 4.

### 4.2. Short series due to large clusters

The second possibility to investigate the spectra observed is to study the relatively short series that can be found over the whole frequency range studied. Several examples will be shown, with measured data and derived results in Tables 2 and 3. In Fig. 10, a typical form of band with narrow-spaced lines of alternating intensity is shown, as the small signal peaks below the three dominating peaks due to $K_{19}$ at $n = 4$ described above. A similar enlarged spectrum is shown in Fig. 11. Note the very apparent alternation in intensity. The spacing of $2B = 32.1$ kHz between the lines agrees well with rotational transitions in cluster $K_{91}$ ($n = 5$). Several other bands with regular spacings of this type are detected repeatedly over a large frequency range from 7 to 80 MHz, with spacings varying between 15.8 and 93.5 kHz. One example is shown in Fig. 12, with a spacing of $2B = 28.9$ kHz. This spacing corresponds accurately to a cluster $K_{37}$ in excitation level $n = 8$. One common feature of all these bands is that they do not extend far towards low frequencies, possibly since they are formed from states with high $J$ values in clusters with large values of $W$. This means that the angular momenta couple as in Fig. 2. The effective $J$ values found are in the range of 500 – 1000, which is considerably smaller than $J_{max}$ corresponding to the peak of a thermal equilibrium distribution. Thus, these bands may be due to pure rotational transitions in angular momentum cases like that depicted in Fig. 2. Another possibility would be that such a band is on top of another transition, like an electron spin-flip process. This is not well studied since the possible center of each band is not easily determined due to lack of a clear structure indicating the center. This indicates anyway that a different process like a spin-flip process is not likely to be involved. Since relatively few lines are observed in many of these bands, the effective cluster overall rotational temperature is rather low. This may be due to the non-thermal population process via stimulated emission in the IR and the (see Fig. 1). Almost all of these bands have spacings that agree very well with rotational transitions in ordinary six-fold symmetric K RM clusters, as summarized in Table 2. The two exceptions are described below. The corresponding dimensional ratios $d/r_n$ are shown in Table 3. It is obvious that important trends exist in these values in the table. Note that the values of $d/r_n$ in the table agree within approximately 1%, thus the agreement with theory is highly significant.

### 4.3. Intermediate series due to spin transitions

Of all relatively narrowly spaced series observed repeatedly, there are only two that do not agree accurately, within 2%, with rotational spacings of the type corresponding to overall rotation of RM clusters. They are both quite apparent and recurring. They have spacings of 51.3 and 127.5 kHz, and tend to have slightly broader lines than the rotational series described above. Examples are shown in Figs. 13 and 14, from which it can be observed that the lines are indeed broader. The width is of the order of 10 kHz. The Lorentzian shape of the peaks may indicate that this width is the natural linewidth. Then, the lifetime of these states is of the order of a few μs, much shorter than for the rotational transitions observed to have sharper lines. This may mean that the nuclei (core ions) are influenced by "collisions" through the electron orbiting motion, with its vibrational variation due to thermal motion. Another more likely possibility is that the orbiting electronic spin, at an angle to the magnetic field from the orbit, influences the magnetic field at the nucleus periodically and thus gives a variation in the field strength. This gives the broadening of the lines.

There is no apparent intensity alternation of the lines in these series. If they are due to spin-flips of any type, their spacing will vary directly with the magnetic field strength, which is given in Eq. (6) for the nucleus position. It is directly found that 127.5 kHz × $5^5$ = 3.984×$10^8$ Hz, and that 51.3 kHz × $6^5$ = 3.989×$10^8$ Hz: these values agree within ± 1×$10^{-3}$. Cf. the formula for the magnetic field strength in Eq. (6). Thus, it is highly likely that these two spacings are due to the magnetic field via spin transitions in clusters with $n$ = 5 and 6 respectively. In general, electronic spin-flips give much too large energies, while nuclear spins are more likely. The values calculated from Eq. (9) were 42.34 kHz for $n$ = 5 and 17.01 kHz for $n$ = 6. Multiplying this with a factor of 3, one finds 127.0 and 51.0 kHz respectively. These values are only 0.4 – 0.6% lower than the measured values, which means a very accurate determination of the spin transitions in view of the unknown cluster sizes for these values and thus less well-known bond distances. The only not understood quantity is the multiplicative factor of 3: it may be due to either a flip from $m_I$ = 3/2 to -3/2, or to the simultaneous flip of three neighbouring nuclei with $\Delta m_I$ = -1. See further the discussion.

## 5. Discussion

### 5.1. Emission from inverted distributions

The evidence that the emission observed is from inverted distributions via stimulated emission is compelling. If the distributions were thermal, practically no signal would be observed at all. The routine use of the RM laser shows that efficient stimulated emission takes place in the IR, in the range between 800 and 16000 nm [8-10]. The extreme self-cooling observed in many other experiments with RM also proves that stimulated emission processes occur easily in the thermal frequency range, at wavelengths longer than 1 µm [3,4]. Thus, it must be expected that stimulated emission will also exist in the RF range. The experiments described here give ample proof on this point: as described in the theoretical section, the signal intensity is many orders of magnitude above the limit that theoretically indicates stimulated emission. While the long series of rotational transitions in principle (apart from the signal strength) could be due to a thermal distribution over the rotational states, the short rotational series observed would probably not be possible from a thermal distribution.

In this respect, it is of course necessary to distinguish between thermal excitation processes, where thermal energy in a non-equlibrium situation is used to excite long-lived states that may radiate by stimulated emission (an effect observed here), and thermal distributions, that can not give rise to stimulated emission since no inversion exists in a thermal (equilibrium) distribution.

**5.2. Bond lengths**

The different bond lengths found from the rotational transitions for clusters with various $n$ and $N$ are given in Table 3 as the corresponding dimensional ratio $d/r_n$. The trends with $n$ and $N$ are quite clear, accepting the uncertainty from the data for the large clusters due to the short series of lines and the apparent line-broadening due to the unresolved different isotopic forms of the clusters. Overall, the agreement with predictions is excellent. It will be assumed for the present discussion that the electron orbiting distances $r_n$ are exactly the theoretically expected values, and thus the variation observed in $d/r_n$ is only due to variation in $d$.

It is expected that the classically predicted value of 2.9 is more accurate the larger the value of $n$ is, since this means that the classical limit is approached. This is also found: of course, the numerical agreement at 2.90 is fortuitous since the precision of the theoretical calculation [6] was not that good. That the value $d/r_n$ is smaller for lower $n$ means probably that the orbiting electrons are not so close to the plane of the cluster.

The other trend observed in Table 3 is that $d$ decreases with increasing $N$, i.e. the bond lengths become smaller for larger clusters. This is expected since the larger clusters will have more atoms surrounded by neigbours and next-neighbours etc., thus being in a more continuous phase. This means that less edge effects will exist, and the edge effects are likely to give larger bond distances due to the open space outside the outermost atom, where the electron density will leak out. Thus, smaller bond lengths in larger clusters are expected.

**5.3. Magnetic dipole transitions**

One important aspect of the results presented is the mechanism by which the non-polar clusters can radiate so intensely under rotation. The radiation energy is due to thermal excitation, and it is high above any continuous thermal background. Stimulated emission is important as described in the theoretical section, similar to the case for IR emission in the RM laser [8-10]. The most important aspect of the emission is related to the mechanism of the transitions. The emission process may be due to electric or magnetic dipoles, the second possibility being a very appealing and likely alternative, since the RM clusters are very strong magnetic dipoles and the power radiated by any multipole will depend on the square of the multipole moments [30]. The magnetic dipole moment of the cluster with $N = 19$ and $n = 4$ is $m = 3 \times 19\ \mu_B = 57\ \mu_B$ from the electron orbital motion [21]. For $n = 6$, $m = 95\ \mu_B$. Thus, the power radiated by the the magnetic dipole in an RM cluster will be $10^4$ times larger than for a typical molecule. The magnetic dipole transition moment has not been computed due to its complexity: the dipoles in the clusters are due to spins, nuclear spins and electronic orbital motion, while the wavefunctions involved are the inseparable $\psi_V \psi_E \psi_R$.

The requirement for an electric dipole transition in the present case is that the cluster rotation is coupled to other degrees of freedom that do give a varying dipole moment. In the theoretical section, it is stated that normal vibrational motion does not seem to exist in RM, mainly due to the break-down of the Born-Oppenheimer approximation which means that the electronic and nuclear motions can not be separated: indeed, vibrations are part of the electronic orbiting motion. Thus, vibrations alone are highly unlikely to give the varying dipole moment. In the case of the RM clusters, the motion of the high-$l$ orbiting electrons gives a magnetic dipole that may couple to the electron spins and also to the nuclear spins of the K atoms. However, the energies corresponding to the orbital motion of the RM electrons at typical $n$ values for RM in the laboratory are much too large to give series of lines in the RF range, as described in the theoretical section. Spin flips of the electrons could possibly be observed, and at $n = 6$ the frequency for a spin flip is close to 24 MHz, thus much larger than the observed spacings. Thus, no mechanism appears to exist by which cluster rotation is coupled to a process giving an electric dipole transition in the spectral range studied. It is concluded that the cause of the emission is the very large magnetic dipole moment of the clusters.

**5.4. Angular momentum couplings**

Other magnetic effects also appear in the spectra, as described in the section of the nuclear spin transitions above. As can be seen in Fig. 8, the series for $N = 19$ clusters at $n = 4$ gives half-integer $J$ numbers. This is not unexpected, since both $S$ and $I$ may be half-integer, corresponding to 19 electron spins with $s = ½$ and 19 nuclear spins with $I = 3/2$. According to the general coupling scheme in Fig. 2, $J$ will be half-integer if $S$ (or $I$, not shown in the figure) is half-integer. Thus, this is expected. The only factor to investigate is why such low values of $J$ are observed. The lowest value of the $J$ for the rotation is 13.5 in Fig. 8, but values down to $J = 9.5$ have been observed in other runs. The coupling situation in Fig. 3 is then probable, where $J$ may even be smaller than $S$ and definitely smaller than $W$. In the case of clusters with $n = 5$ and 6, integer $J$ quantum numbers are however found. As seen in Table 1, the magnetic field strength decreases a factor of approximately three in each step from $n = 4$ going to 5 and 6. Thus, a direct coupling between $\underline{s}$ and $\underline{I}$ for each atom or between $\underline{S}$ and $\underline{I}$ may be possible at $n > 4$, which would give integer values of $J$. One may compare this to a Zeeman coupling at low magnetic field, and a Paschen-Back coupling of each vector to the field direction at strong magnetic field, i.e. with $n = 4$. That the coupling scheme is changing at $n = 4$ is supported by other experiments that show two simultaneous series for $n = 4$, both with integer and half-integer quantum number values for $K_{19}$. Such a change in behaviour in different runs may certainly be due to the actual form of the clusters formed at the specific densities and temperatures, which will change the spectrum of couplings in between the two cases shown in Figs. 3 and 4.

## 5.5. Centrifugal distortion

The long series of lines do not show any departure from a linear behaviour, thus they show no centrifugal distortion. This is expected, since the rotation of the clusters is extremely slow. At a typical value $R = 20$, the rotation of the $K_{19}$ cluster is slow, close to $2 \times 10^6$ s$^{-1}$, and of very low energy. Note that the thermal maximum (not really existing due to the stimulated processes) would be at $L = 1000$ even at $T = 100$ K due to the extremely small $B$ value. Since the vibrations in the RM clusters can not be separated from the electronic motion, a direct calculation of the centrifugal distortion constant $D$ from the $B$ value is not possible. However, since $D/B$ varies as $B^2$, the very small values of $B$ (< 1 MHz instead of > 1 GHz for small molecules), $D$ will be less significant in the present case by a factor of $> 10^6$. Further, one may realize that the centrifugal force on a K atom in the $K_{19}$ ($n = 4$) cluster at $R = 20$ is of the order of $5 \times 10^{-19}$ N, thus very low. This value should be compared to the centrifugal force of the order of $7 \times 10^{-10}$ N for a small molecule like HCl where the distorsion constant is 12 MHz. Thus, the stretching force is $10^9$ times smaller in the $K_{19}$ cluster. These two factors determine the value of $D$, and both factors indicate a value of $D$ many orders of magnitude smaller than in small molecules. Thus, no departure from the linear behaviour should be observed in the experiments.

Another factor that may prevent the observation of any centrifugal stretching is the fact that the coupling scheme in Fig. 3 is valid for the longest series studied with best precision, for $K_{19}$ ($n = 4$). This means that different $J$ values do not have to correspond to different $R$ values, i.e. a higher $J$ value does not mean a faster overall rotation of the cluster since $J$ will also depend directly on S and probably also on $I$ (not included in detail here due to its complexity). Of course, this does not mean that a rotational transition does not imply a change in $R$, especially in $K_b$, but the different $J$ numbers observed may not differ in $R$ but only in $S$ (and $I$). In fact, a slower overall rotation may even correspond to a higher $J$ value in a subclass of clusters. Thus, it would be safest to use large clusters in high $n$ levels, where the schemes in Figs. 2 and 4 may be applicable, to investigate the centrifugal stretching.

## 5.6. Intensity periodicities

The short rotational series found for the larger clusters display an alternating intensity of the peaks, as seen in Figs. 10 - 12. Such a variation is expected due to the nuclear spin statistical weights of the K nuclei with nuclear spin I = 3/2. This means that they are Fermions and that the total wavefunction of the cluster should be antisymmetric under exchange of two of the nuclei. The clusters belong to point group $C_{6h}$, for which the rotational point group is $C_6$. The non-separable electronic and vibrational wavefunction $y_E y_V$ must be used since these degrees of freedom can not be separated in RM clusters, as described in the theoretical section. This function is not totally symmetric, but probably belongs to symmetry species $A_2$. Thus, the function $y_R y_S$ must be type $B_2$ to give the total wave function the type $B_1$ which is the antisymmetric type in this point group [31]. The *K* values of the clusters and their relations to *J* are unspecified, thus only the odd/even variation in *J* with *K* = 0 will give any intensity variation. If we use the same spin functions as for hydrogen in $C_6H_6$ in Ref. [31], a variation with weights 7 (*J* odd): 3 (*J* even) is found. This treatment is clearly not correct even for a cluster $K_7$ since the spin for the K nucleus has four different states. Despite this, the high resolution spectrum in Fig. 12 gives an average of 2.47 ± 0.48 for the ratio between the peaks, in approximate agreement with the theoretical value 7/3 = 2.33. Thus, this approximate treatment gives approximate agreement, and better data for specific clusters with *K* = 0 are needed if any more information should be found from the intensity alternations.

The most interesting result from this discussion may be that the main variation is in every second peak, and that this variation with *J* should exist only for *K* = 0. Such states are most probable with the coupling scheme shown in Fig. 4, where *K* close to 0 means that *R* is perpendicular to $\underline{W}$. The intensity alternations are only detected for large clusters with $N = 37$, 61 and 91, that are not of the minimal interaction type. This is in agreement with the descriptions of the angular momentum coupling schemes, since for the small clusters where low $J$ states can be observed, $K$ is probably quite large.

5.7. **Nuclear spin transitions**

The lines attributed to nuclear spin transitions are different from the rotational series of transitions, both concerning the absence of intensity alternations and the widths of the peaks. The numerical agreement with the theoretical calculations of the line spacings is very good, accepting the fact that the observed transitions are three times larger than the theoretical value. This may indicate that three nuclei combine in the transitions, each changing $\Delta m_I = +1$ of -1, or that the spin changes by $\Delta m_I = +3$ or -3, e.g. from 3/2 to -3/2 in one transition. (Other possibilities exist, see below). The appearance of the spectra, with many equally spaced peaks, indicates anyway that the spins are strongly coupled in the RM at $n = 5$ and 6, corresponding to spacings of 127.0 and 51.0 kHz respectively. The magnetic field from the orbiting motion of the electrons is quite strong at the nuclei, as described in the theoretical section. This means that each nuclei can be in four different states with values $m_I = 3/2, 1/2, -1/2$ and -3/2. The energy levels due to the magnetic field are given in Eq. (9). As described in the theoretical section, the quantum number $F$ from $\underline{F} = \underline{J} + \underline{I}$ determines the energy level. In the same way as for an overall rotation, the energy levels due to different $F$ are spaced unevenly giving many different lines separated by $a$, the magnetic dipole interaction constant [27]. Each such line is designated by a number $F$, e.g. for the lower level in the transition.

The nuclear spin series of transitions are more localized in frequency than the rotational series are. The series with spacing of 51.0 kHz (in RM with $n = 6$) is observed at 43 - 48 MHz, thus higher than the electron spin transition in RM ($n = 6$) which should appear at 24 MHz, slightly dependent on the position in the cluster. A coupling to a transition in $S$ is likely in this case, since otherwise the $F$ value would be around 2600. This is too high to be realistic at least for the cluster sizes studied here: the total $L$ for a cluster with $N = 91$ and $n = 6$ is only 546. Thus, a coupling to a spin or other type of transition is probable in this case; otherwise, one has to imply the existence of even larger clusters not observed directly in previous experiments.

The nuclear spin series with spacing 127.0 kHz ($n = 5$) is observed between 24 and 32 MHz, where no electron spin transition at $n = 5$ is possible. This would indicate an $F$ value of the order of 700, while total $L$ for a cluster with $N = 91$ and $n = 5$ is 455. With $\underline{S}$ and $\underline{R}$ included, the observed range may be possible. However, as seen in Fig. 14 two such series are possible simultaneously. This indicates that other degrees of freedom like overall rotation are involved in the $F$ transitions that are observed for $n = 5$.

Due to the exact agreement between experiment and theory (within 0.6%), it is concluded that the transitions observed are $\Delta F = -3$. It might be speculated that this selection rule is due to the magnetic dipole nature of the transitions. However, no such effect is observed for the transitions that only change the $J$ quantum number. The most likely explanation at present is that there exists an intensity alternation in the lines, with every third line much stronger. This could be due to nuclear spin coupled to rotation of the cluster around the $I_C$ axis, but the exact mechanism is not known.

## 6. Conclusions

Long series of rotational emission lines for planar RM clusters $K_{19}$ at excitation levels $n = 4$, 5 and 6 are now observed at radio frequencies in the range 7 – 90 MHz, and short series for larger clusters $K_{37}$, $K_{61}$ and $K_{91}$. The transitions are due to the large magnetic dipoles of the rotating clusters. In total, nine different cluster types are reported, with bond lengths in agreement with theory within ±1.5%. Nice trends are observed with excitation level and cluster size. The dimensional ratio $d/r_n$ is now determined in $K_{19}$ ($n = 4$) with a precision of $1 \times 10^{-4}$. The complex angular momentum couplings in the clusters are understood. The nuclear spins are observed in series of apparent transitions $\Delta F = -3$ in the magnetic field from the orbiting motion at $n = 5$ and 6. The agreement with theory is within 0.6%.

**Acknowledgement**

I am very grateful to Daniel Berg, ETA Chalmers Student Union, and Kjell Brantervik, Chalmers Lindholmen, for the loan of the spectrum analyzers, and to Robert Svensson for arranging these loans.

# References

[1] J. Wang, L. Holmlid, Chem. Phys. Lett. 295 (1998) 500.

[2] J. Wang, L. Holmlid, Chem. Phys. 277 (2002) 201.

[3] S. Badiei, L. Holmlid, Int. J. Mass Spectrom. 220 (2002) 127.

[4] S. Badiei, L. Holmlid, Chem. Phys. 282 (2002) 137.

[5] S. Badiei, L. Holmlid, Phys. Lett. A 327 (2004) 186.

[6] L. Holmlid, Chem. Phys. 237 (1998) 11.

[7] S. Badiei, L. Holmlid, J. Phys.: Condens. Matter 16 (2004) 7017.

[8] S. Badiei, L. Holmlid, Chem. Phys. Lett. 376 (2003) 812.

[9] L. Holmlid, J. Phys. B: At. Mol. Opt. Phys. 37 (2004) 357.

[10] S. Badiei, L. Holmlid, Appl. Phys. B 81 (2005) 549.

[11] É. A. Manykin, M. I. Ozhovan, P. P. Poluéktov, Sov. Tech. Phys. Lett. 6 (1980) 95 [Pis´ma Zh. Tekh. Fiz. (USSR) 6 (1980) 218].

[12] É. A. Manykin, M. I. Ozhovan, P. P. Poluéktov, Sov. Phys. JETP 75 (1992) 440.

[13] É. A. Manykin, M. I. Ozhovan, P. P. Poluéktov, Sov. Phys. JETP 75 (1992) 602.

[14] R. Svensson, L. Holmlid, L. Lundgren, J. Appl. Phys. 70 (1991) 1489.

[15] V. I. Yarygin, V. N. Sidel´nikov, I. I. Kasikov, V. S. Mironov, S. M. Tulin, JETP Lett. 77 (2003) 280.

[16] G. I. Mishin, Doklady Akademii Nauk 372 (2000) 612.

[17] J. J. Gilman, Appl. Phys. Lett. 83 (2003) 2283.

[18] A. Kotarba, G. Adamski, Z. Sojka, S. Witkowski, G. Djega-Mariadassou Studies in Surface Science and Catalysis (International Congress on Catalysis, 2000, Pt. A) 130A (2000) 485.

[19] A. Kotarba, J. Dmytrzyk, U. Narkiewicz, A. Baranski, React. Kin. Catal. Lett. 74 (2001) 143.

[20] A. Kotarba, A. Baranski, S. Hodorowicz, J. Sokolowski, A. Szytula, L. Holmlid, Catal. 67 (2000) 129.

[21] S. Badiei, L. Holmlid, Mon. Not. R. Astron. Soc. 335 (2002) L94.

[22] L. Holmlid, Phys. Rev. A 63 (2001) 013817.

[23] L. Holmlid, Appl. Phys. B 79 (2004) 871.

[24] L. Holmlid, Astrophys. Space Sci. 291 (2004) 99.

[25] J. M. Hollas, "High resolution spectroscopy", 2$^{nd}$ ed. Wiley, Chichester, 1998.

[26] J. Wang, K. Engvall, L. Holmlid, J. Chem. Phys. 110 (1999) 1212.

[27] S. Svanberg, "Atomic and Molecular Spectroscopy", 2$^{nd}$ ed., Springer, Berlin 1997.

[28] L. Holmlid, J. Phys. Chem. A 108 (2004) 11285.

[29] G. Meima, P. G. Menon, Appl. Catal. A 212 (2001) 239.

[30] J. D. Jackson, "Classical electrodynamics", Wiley, New York, 1966.

[31] E. B. Wilson, J. Chem. Phys. 3 (1935) 276.

Table 1

Electron orbit radius $r_n$, approximate interatomic distances $d = 2.9\, r_n$, and magnetic field strength $B_n$ at the position of the nuclei according to Eq. (6), as a function of excitation level (principal quantum number) $n$.

| $n =$ | 4 | 5 | 6 | 7 | 8 |
|---|---|---|---|---|---|
| $r_n$ (nm) | 0.84668 | 1.3229 | 1.9050 | 2.59 | 3.39 |
| $d$ (nm) | 2.455 | 3.837 | 5.525 | 7.52 | 9.82 |
| $B_n$ (mT) | 10.3 | 3.39 | 1.36 | 0.630 | 0.323 |

Table 2.

Values of 2$B$ in kHz, with $B$ the rotational constant, experimentally found and theoretical (in parantheses), as a function of the magic number $N$ of the cluster K$_N$. $n$ is the excitation level in the cluster.

| $n =$<br>$N =$ | 4 | 5 | 6 | 7 | 8 |
|---|---|---|---|---|---|
| 19 | 1858.5 ± 0.2 | 744.5 ± 0.6 | 353.7 ± 0.3 | (191) | (112) |
| 37 | (462) | (189) | 93.5 ± 0.5 | 49.3 ± 0.1 | 28.9 ± 0.1 |
| 61 | (169) | 72.3 ± 0.2 | (33.3) | (18.0) | (10.5) |
| 91 | (75.4) | 32.1 ± 0.1 | 15.80 ± 0.05 | (8.0) | (4.7) |

Table 3.

Variation of the dimensional ratio $d/r_n$ i.e. the ratio between the experimental interatomic distance and the theoretical electron orbiting radius, with excitation level (principal quantum number) $n$ and with magic number $N$ of the clusters $K_N$. The average value of $d/r_n$ is 2.865 ± 0.032. Cf Table 2.

| $n =$<br>$N =$ | 4 | 5 | 6 | 7 | 8 |
|---|---|---|---|---|---|
| 19 | 2.8470 | 2.879 | 2.900(4) | | |
| 37 | | | 2.866 | 2.900 | 2.90 |
| 61 | | 2.83 | | | |
| 91 | | 2.84 | 2.82 | | |

Table 4.

Experimentally determined rotational parameters and bond distances in the K$_{19}$ clusters.

| $n$ | $B$ (MHz) | $I_B$ (kg m$^2$) | $d_{theory}$ = 2.9 $r_n$ (nm) | $d$ (exptl) (nm) | $d / r_n$ (exptl) |
|---|---|---|---|---|---|
| 4 | 0.9292 ± 0.0001 | 9.031x10$^{-42}$ | 2.455 | 2.4105 ± 0.0003 | 2.8470 |
| 5 | 0.37225 ± 0.0003 | 2.2544x10$^{-41}$ | 3.837 | 3.8085 ± 0.003 | 2.8788 |
| 6 | 0.17685 ± 0.00015 | 4.745x10$^{-41}$ | 5.525 | 5.525 ± 0.005 | 2.9004 |

**Figure captions**

Fig. 1. The processes that form and deexcite the RM clusters giving the stimulated emission in the RF range. The electronic process giving the IR emission as indicated is described in other studies [9,10].

Fig. 2. The general angular momentum coupling scheme useful for the symmetric rotor clusters studied, with nuclear spin excluded.

Fig. 3. The extreme angular momentum coupling for the symmetric rotor clusters in their minimum interaction state, to the left. No magnetic (or electric) dipole transitions are possible in this case. To the right, a slightly disturbed state of the clusters, in the present study applicable to the case $n = 4$. Note that $J$ is small, much less than $\mathbf{W}$.

Fig. 4. Angular momentum coupling scheme applicable to large clusters in the present study. Due to the weaker magnetic field with $n > 4$, $\underline{l}$ and $\underline{s}$ for each atom are coupled and only $\underline{\mathbf{W}}$ is shown here. $\underline{K_c}$ is small, since $\underline{R}$ is close to the plane of the cluster. This means that the magnetic dipole coupling to the radiation field is strong.

Fig. 5. Emitter from Chamber 1, showing a few dark pieces of catalyst in the bottom of a small Pt foil boat. The catalyst is dark due to coverage of graphite on the surface. The two stainless steel rods carry the current through the Pt foil. The squares on the support are 5 mm wide.

Fig. 6. Emitters of the type used in Chambers 4 and 5. To the right, a fresh unused emitter is shown, with the planar cut surface clearly visible. The Ta foil is folded around the emitter samples. In the center, a dark used emitter is shown, with fresh catalyst samples to the left. The squares on the support are 5 mm wide.

Fig. 7. Radio frequencies detected with analysis bandwidth of 3 kHz at a signal level above approximately 1 µV. Highest displayed signal for each plot is shown at the lefthand vertical axis. A few higher peaks exist in each run. For a description of the types of chambers used, see the text.

Fig. 8. A linear plot of 30 RF frequencies measured in an experiment from Chamber 1. The slope is found to be 1.8583 ± 0.0003 MHz. The zero is found at -0.1410 ± 0.0101 MHz. The ratio of these values gives the zero at $J = 0.076$ which indicates that the $J$ values are half-integer as shown.

Fig. 9. Linear plots of experiments at high emitter temperatures. Fit parameters are for the $n = 5$ line slope 0.7445 ± 0.0006 MHz, zero at -0.1192 ± 0.0450 MHz; and for $n = 6$ line, slope 0.3537 ± 0.0003 MHz with zero at 0.0751 ± 0.0385 MHz. The points indicated have integer $J$ values.

Fig. 10. The typical RF signal at 44-50 MHz in Chamber 3. The three strongest peaks are separated by 1.8585 MHz, belonging to the series corresponding to rotation of $K_{19}$ ($n = 4$). The spacing of the peaks in the band to the left is 32.1 kHz which is due to cluster $K_{91}$ ($n = 5$). Analysis bandwidth 3 kHz. See further the text and Fig. 11.

Fig. 11. A band of rotational transitions due to $K_{91}$ ($n = 5$) with spacing 32.1 kHz. The large peak belongs to the series due to $K_{19}$ ($n = 4$) as shown in Fig. 10. Note the intensity alternations that indicate a rotational series.

Fig. 12. Part of a band with rotational transitions due to $K_{37}$ ($n = 8$) with spacing 28.9 kHz. Observe the intensity alternation.

Fig. 13. A short series of rotational transitions (sharp peaks) corresponding to $K_{37}$ ($n = 7$), almost hidden in a longer series of nuclear spin transition peaks (broad), with a spacing of 51.6 kHz. This spacing corresponds to spin transitions in RM at $n = 6$ with $\Delta F = -3$. The analysis bandwith is 3 kHz.

Fig. 14. Part of a band extending over a total of 4 MHz. Spectrum taken with analysis bandwidth of 300 Hz. Besides two solitary peaks close to 24.2 MHz, two series with broad peaks and spacings 127.5 kHz (nuclear spin transitions at $n = 5$) and one series with sharper peaks with spacing 121.6 kHz, possibly $K_{18}$ ($n = 8$), are observed. Observe the intensity alternations in this last rotational series.

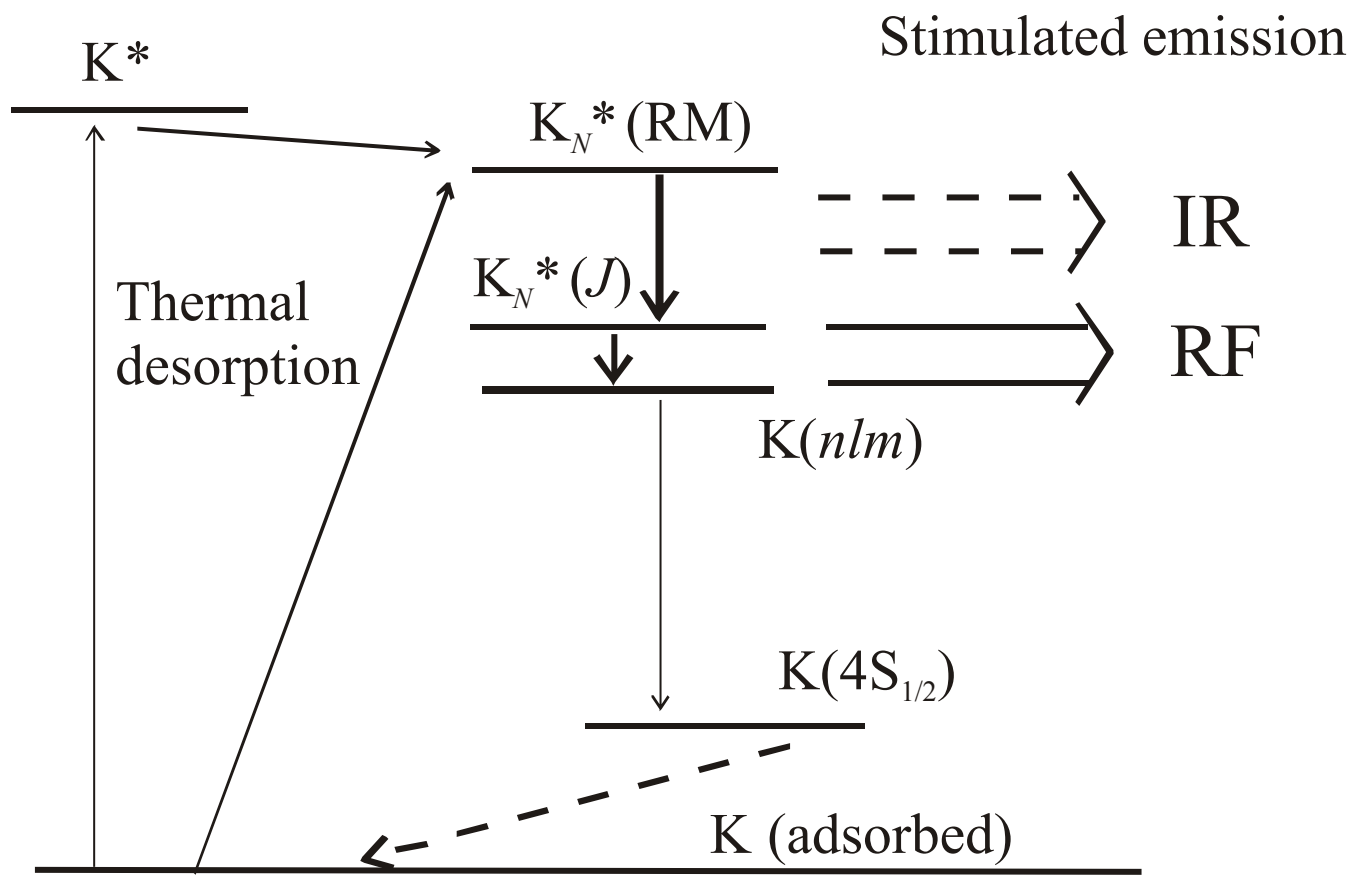

Fig. 1

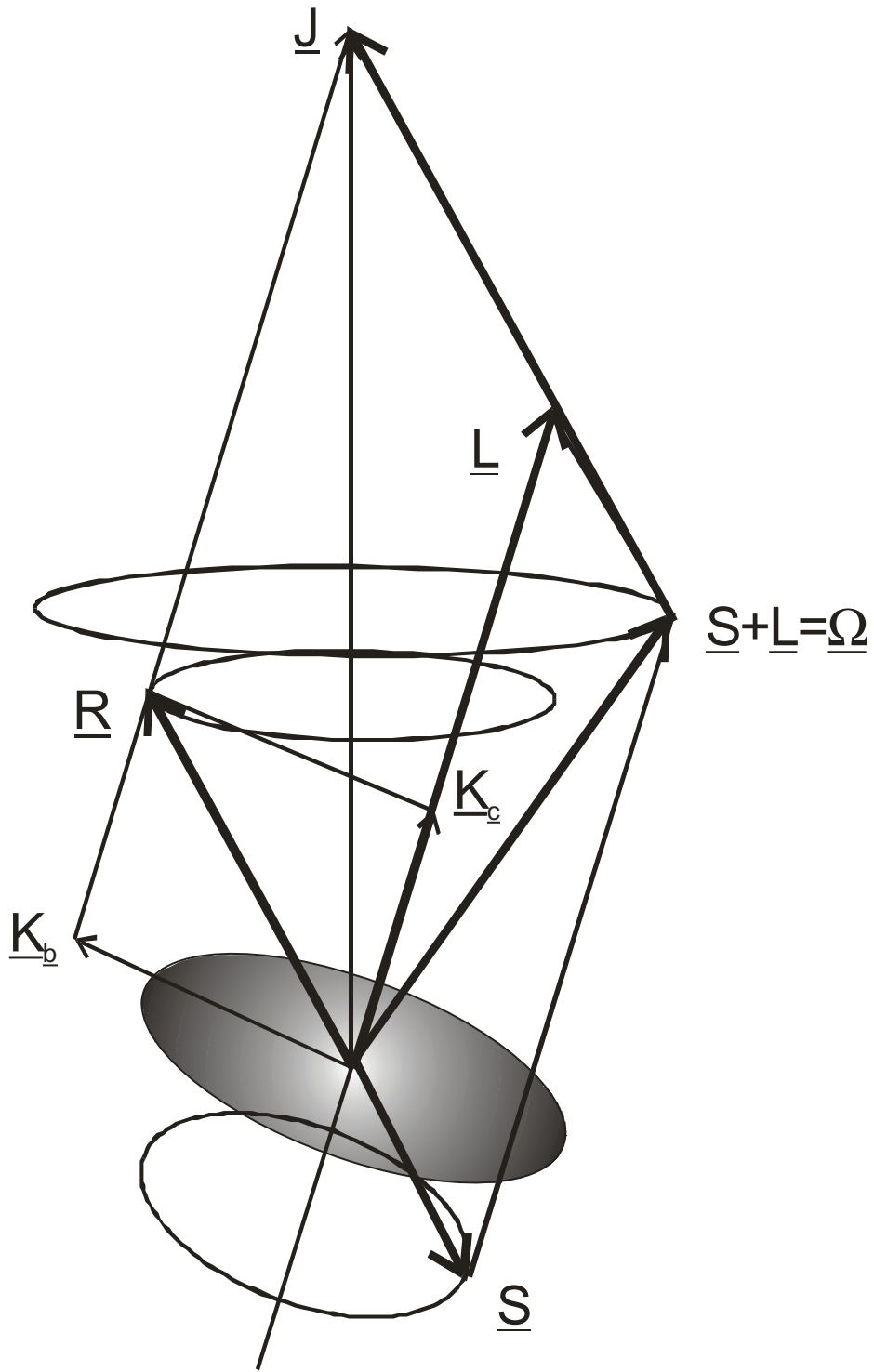

Fig. 2

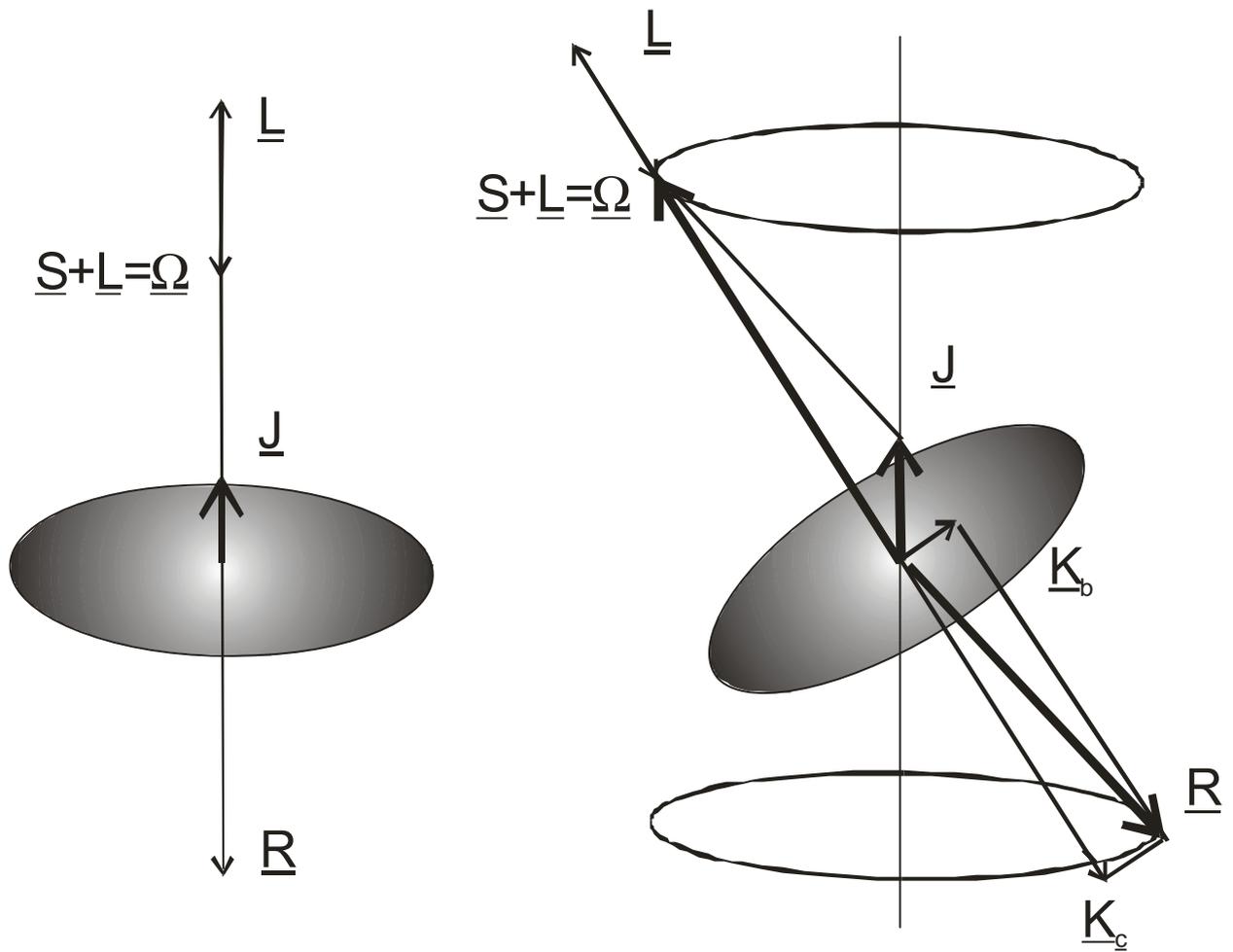

Fig. 3

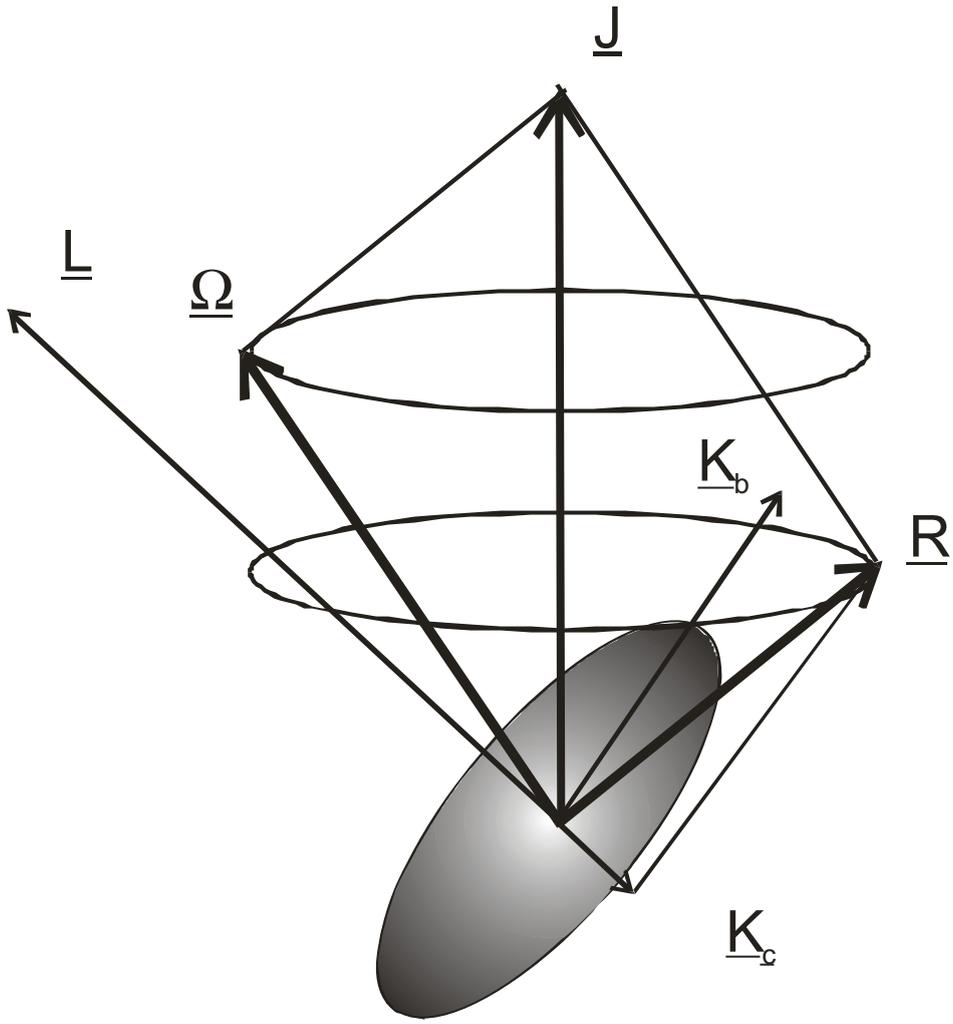

Fig. 4

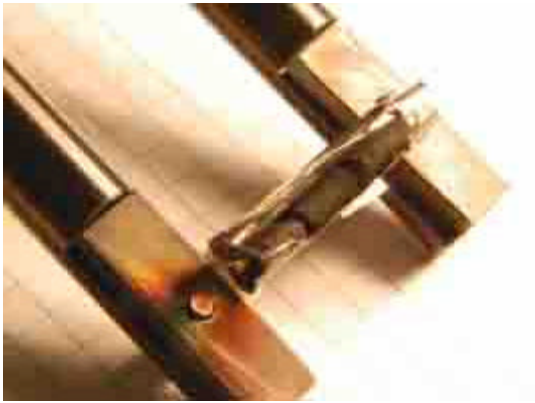

Fig. 5.

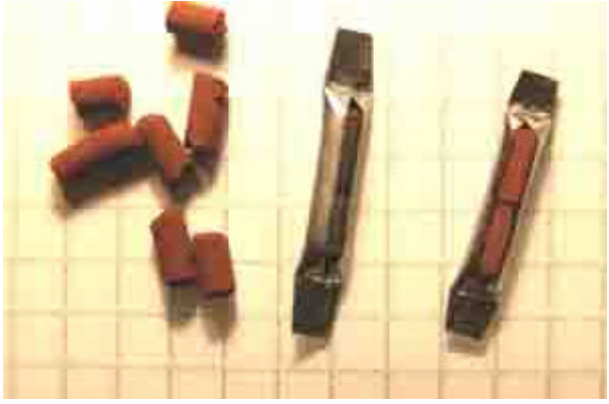

Fig. 6

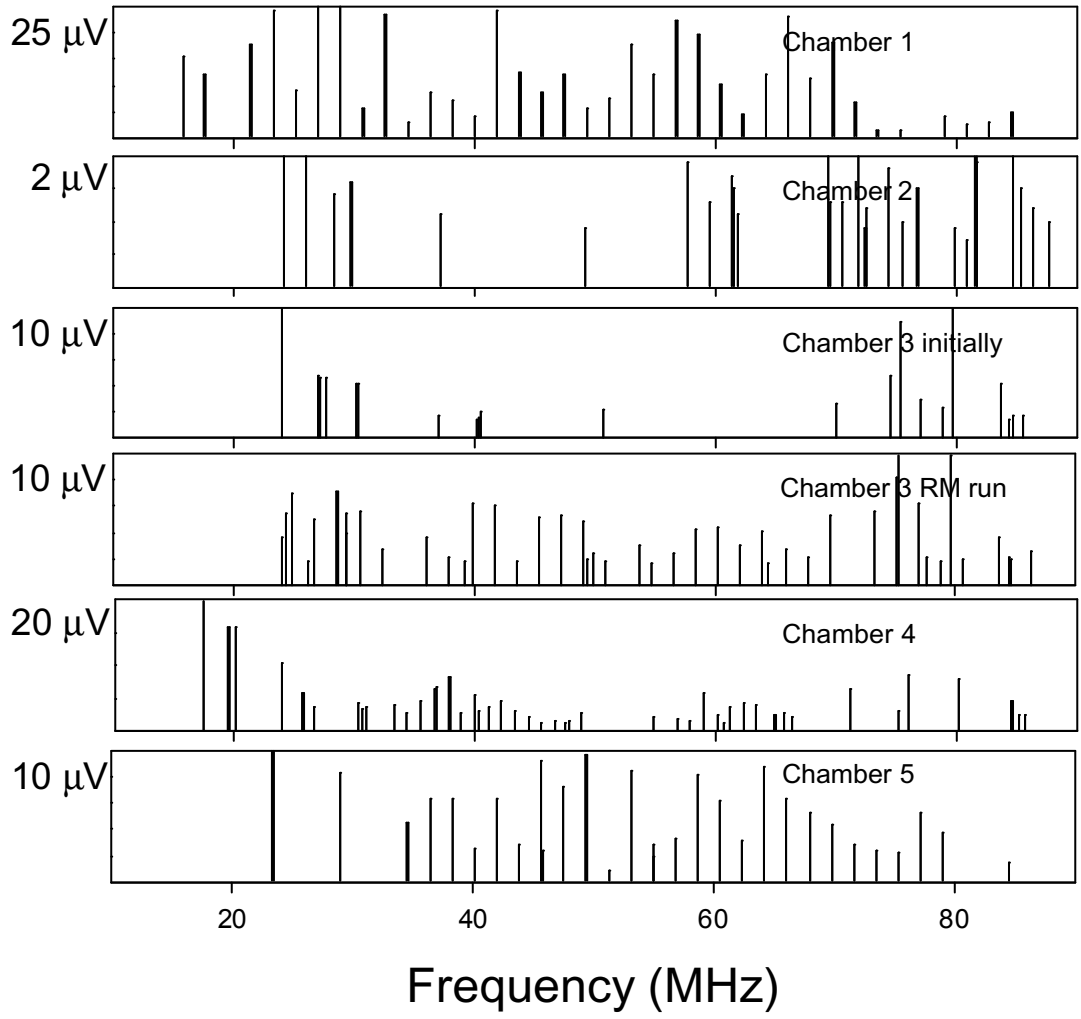

Fig. 7

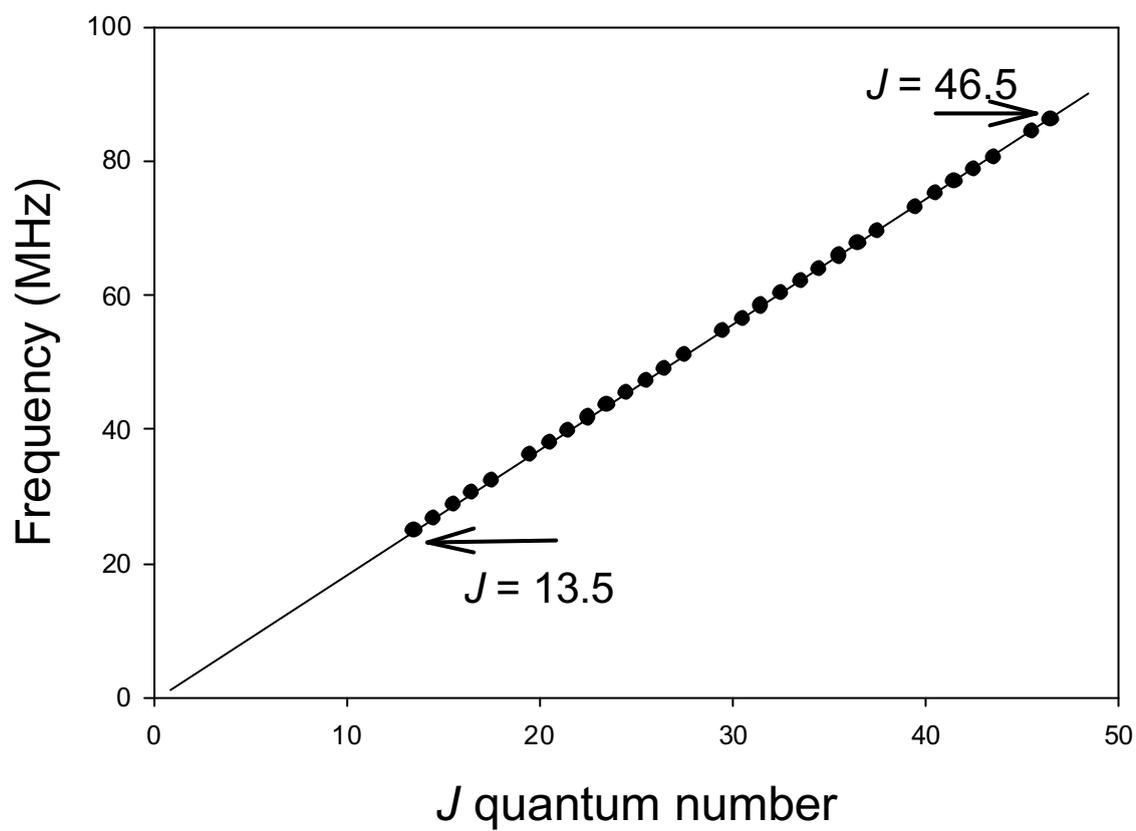

Fig. 8

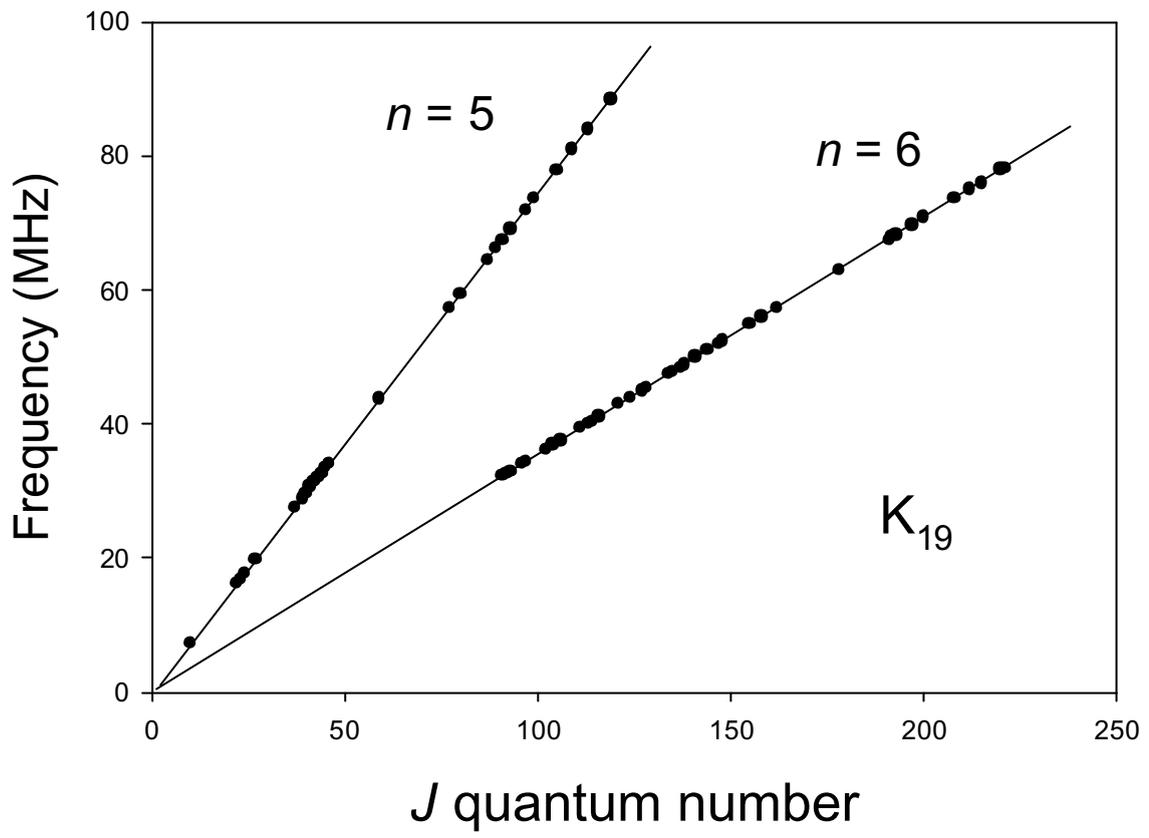

Fig. 9

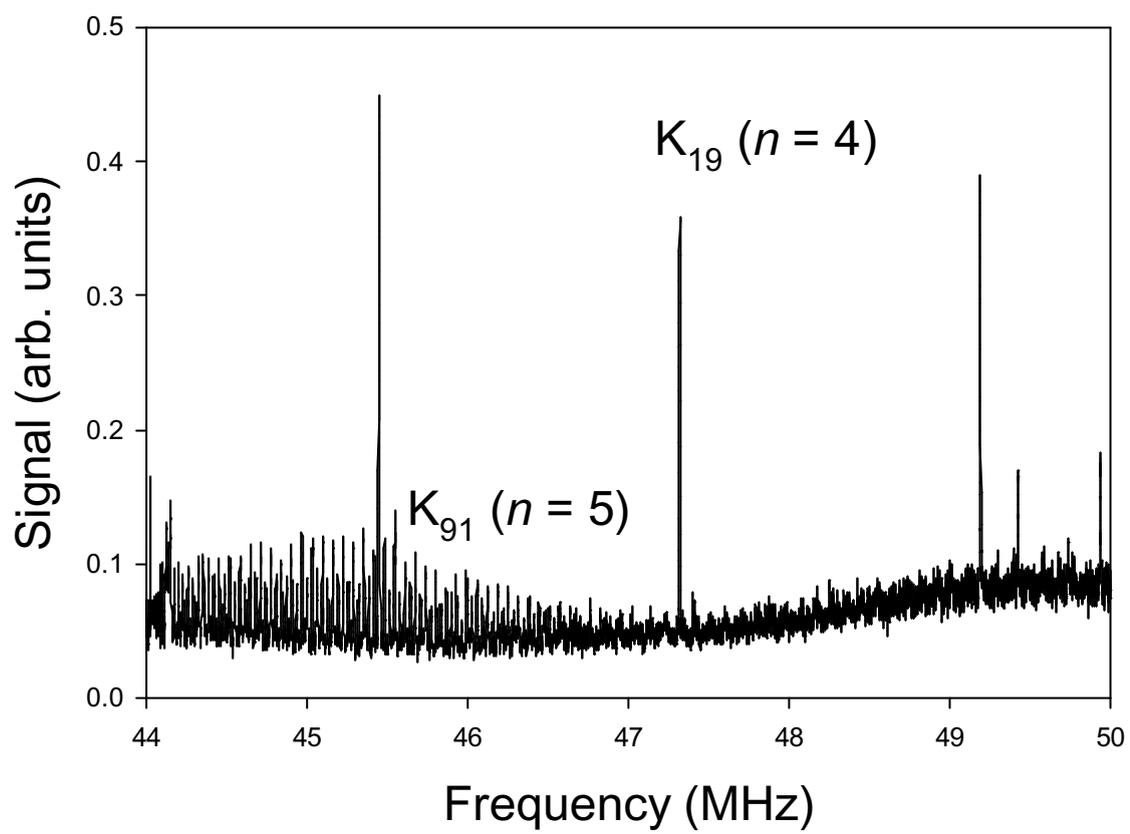

Fig. 10

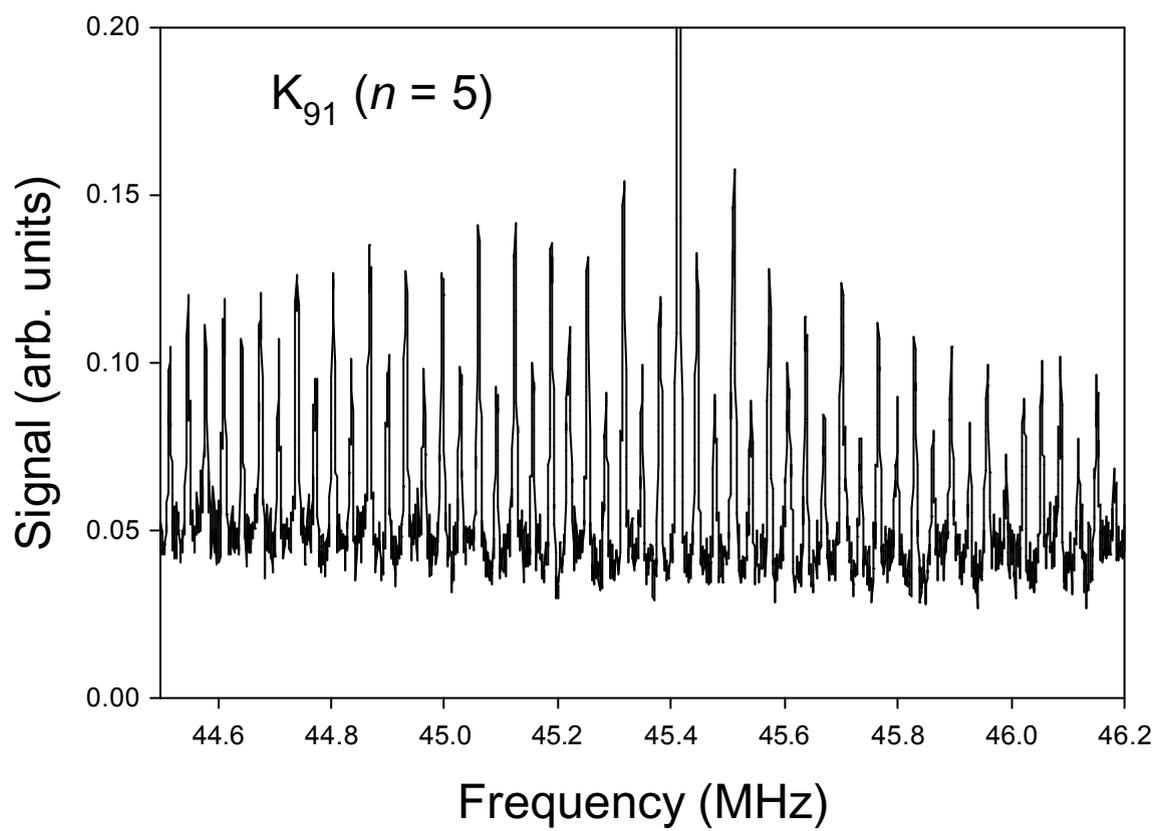

Fig. 11

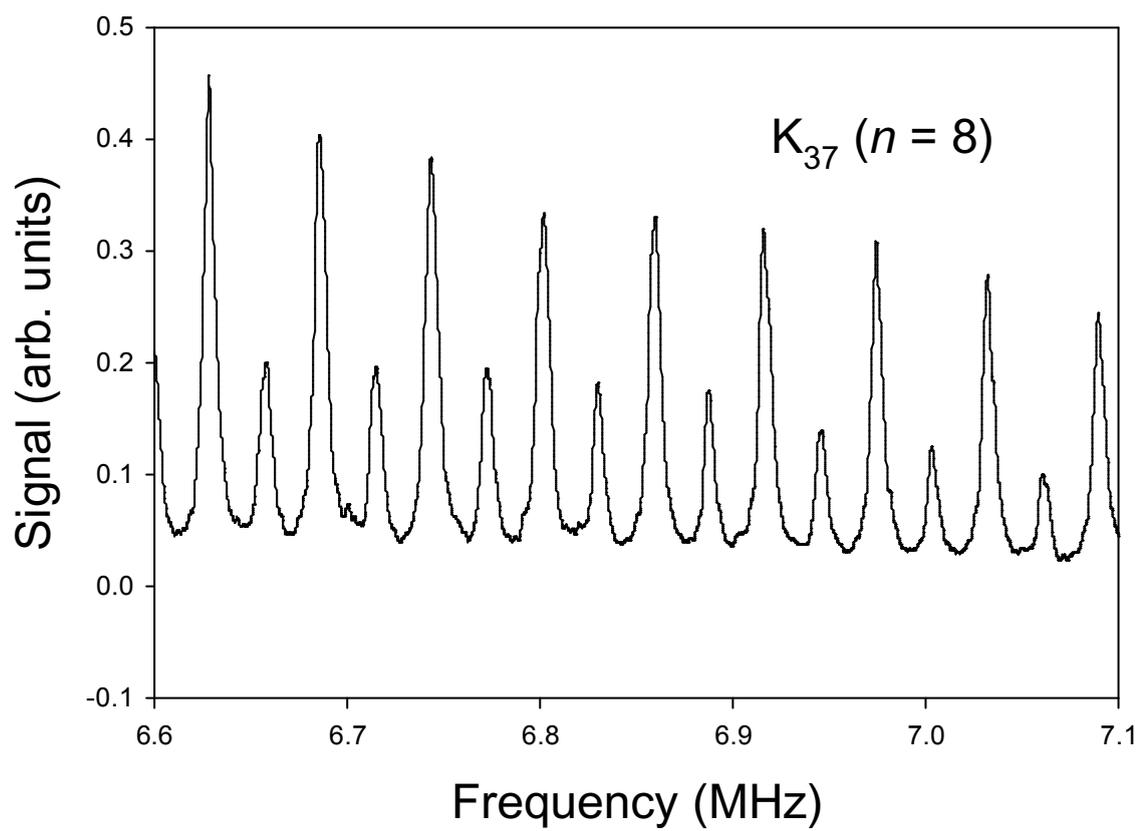

Fig. 12

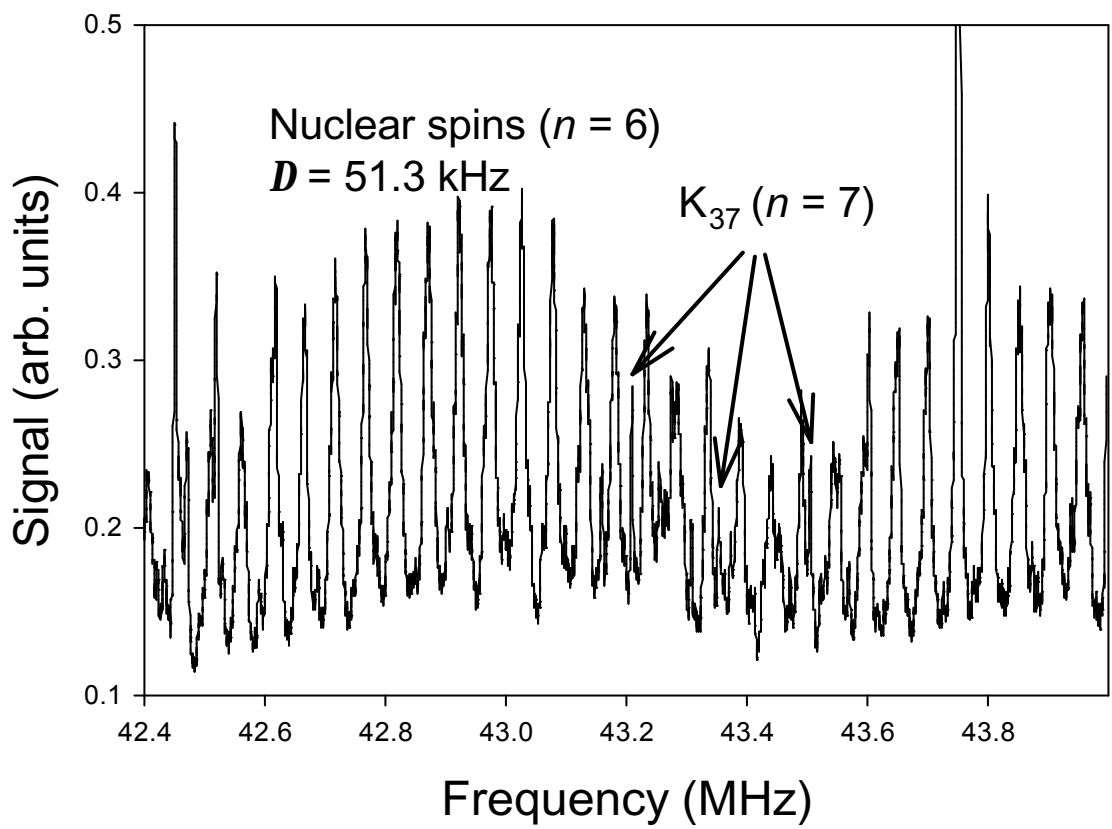

Fig. 13

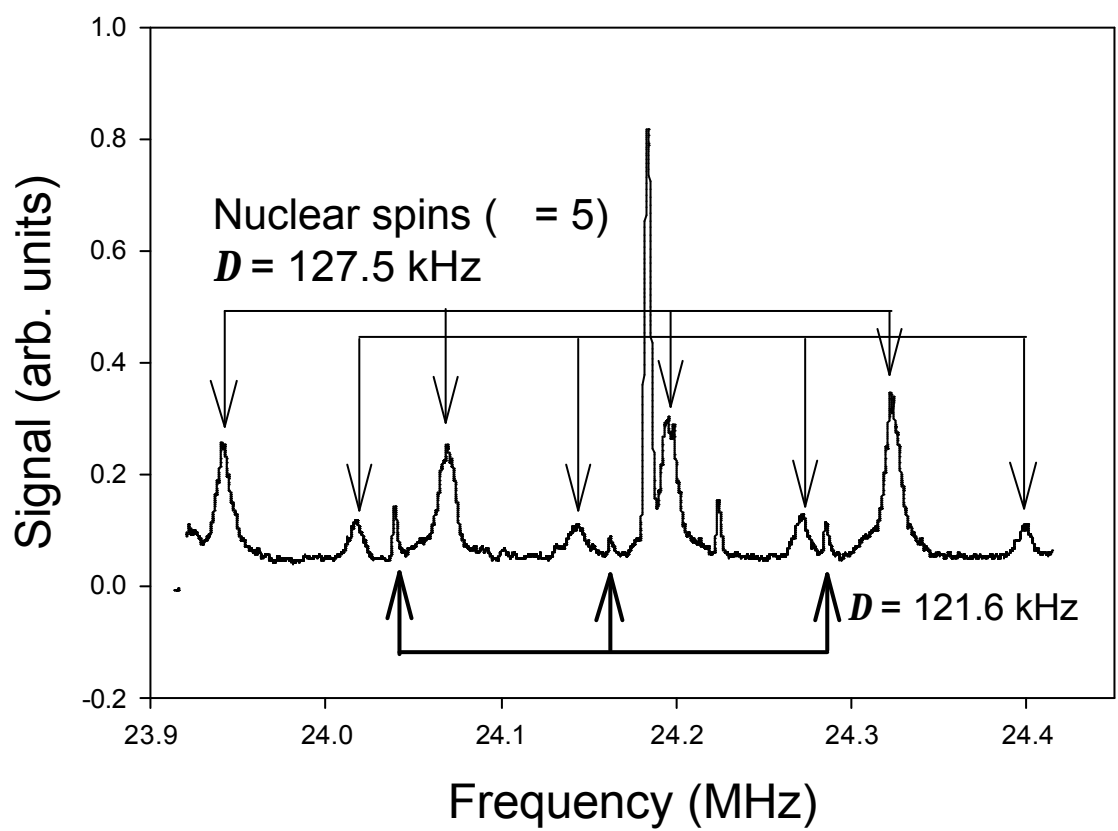

Fig. 14